\documentclass[aps,prc,twocolumn,superscriptaddress,showpacs]{revtex4}

\usepackage{graphicx}
\usepackage{latexsym}
\usepackage{bm}
\usepackage{amsmath}

\usepackage{color}

\newcommand{\comment}[1]{}

\newcommand{\gsim}{\mbox{\raisebox{-0.6ex}{$\stackrel{>}{\sim}$}}\:}
\newcommand{\lsim}{\mbox{\raisebox{-0.6ex}{$\stackrel{<}{\sim}$}}\:}

\begin{document}

\title{
Interplay between soft and hard hadronic components
for identified hadrons
    in relativistic heavy ion collisions
}

\author{Tetsufumi Hirano}
\affiliation{Physics Department, University of Tokyo,
  Tokyo 113-0033, Japan}
\affiliation{RIKEN BNL Research Center,
    Brookhaven National Laboratory, Upton, New York 11973}
\author{Yasushi Nara}
\affiliation{%
Department of Physics, University of Arizona, Tucson, Arizona 85721
}

\date{\today}
 
\begin{abstract}
We investigate the transverse dynamics
in Au+Au collisions at $\sqrt{s_{NN}}=200$ GeV
by emphasis upon the interplay between soft and hard components
through $p_{\mathrm{T}}$ dependences of particle spectra, ratios of yields,
suppression factors, and elliptic flow for identified hadrons.
From hydrodynamics combined with traversing minijets
which go through jet quenching in the hot medium,
we calculate interactions of hard jets
with the soft hydrodynamic components.
It is shown by the explicit dynamical calculations that
 the hydrodynamic radial flow and the jet quenching of hard jets
are the keys to understand the differences among the hadron
spectra for pions, kaons, and protons.
This leads to the natural interpretation for $N_p/N_\pi \sim 1$,
$R_{AA} \gsim 1$ for protons, and 
$v_{2}^{p} > v_{2}^{\pi}$
recently observed
in the intermediate transverse momentum region
at Relativistic Heavy Ion Collider (RHIC).
\end{abstract}


\maketitle

\section{Introduction} 

A vast body of data has already been collected and analyzed
  during past few years at Relativistic Heavy Ion Collider (RHIC)~\cite{QM}
  toward a complete understanding of the dense QCD matter
  which is created in high energy heavy-ion collisions.

At collider experiments, it is well known that
 high $p_{\mathrm{T}}$ perturbative QCD (pQCD) processes become
so large as to observe jet spectra.
One of the most important new physics
  revealed in heavy ion collisions at RHIC energies
 is to study propagation of (mini-)jets in dense QCD matter.
Jet quenching
has been proposed~\cite{Gyulassy:1990ye}
 as a possible signal of deconfined nuclear matter,
 the quark gluon plasma (QGP)
(for a recent review, see Ref.~\cite{Gyulassy:2003mc}).
Over the past years, a lot of work
has been devoted to study the propagation
of jets through QCD matter \cite{Baier,Wiedemann,Zakharov,Levai}.

Recent data at RHIC indicate
  that both the neutral
  pion~\cite{phenix:pi0_130,phenix:pi0_200}
  and the charged hadron
  ~\cite{star:highpt,star:charged,Adcox:2002pe} spectra
  at high $p_{\mathrm{T}}$ in central Au+Au collisions
  are suppressed relative to
  the scaled $pp$ or large centrality spectra
  by the number of binary collisions.
However, protons do not seem to be quenched
 in the moderate $p_{\mathrm{T}}$ range \cite{Adler:2003kg}.
Furthermore,
 the proton yield exceeds
  the pion yield around $p_{\mathrm{T}} \sim 2$-3 GeV/$c$
   which is not seen in elementary hadronic collisions~\cite{Adcox:2002pe}.
The STAR Collaboration
 also shows that $\Lambda/K^0 \sim 1$ at a transverse momentum
of 2-3 GeV/$c$~\cite{STAR:Lambda_K}.
pQCD calculations are successful in describing hadron spectra
in Au+Au collisions as well as $pp$ collisions
by taking account of nuclear effects such as Cronin effect,
nuclear shadowing effect, and energy loss of jets~\cite{VG}.
However, large uncertainly of the proton fragmentation function
makes the understanding of the baryon production mechanism
unclear~\cite{XZhang} even in $pp$ collisions.
On the other hand,
several models have been proposed by considering interplay
between non-perturbative soft physics and pQCD hard physics:
  baryon junction~\cite{Vitev:2002wh,ToporPop:2002gf},
  parton coalescence~\cite{AMPT,Hwa:2002tu,Fries,Greco,Molnar:2003ff},
  medium modification of the string fragmentation~\cite{Casalderrey:2003cf},
and
 a parametrization with hydrodynamic component combined with
 the non-thermal components~\cite{Peitzmann}
in order to explain the anomalous baryon productions
and/or large elliptic flow discovered at RHIC.

It is said that hydrodynamics~\cite{KHHH,Teaney:2000cw,Hirano:2001eu,
HiranoTsuda}
works very well
for explanation of elliptic flow data at RHIC energies,
 in the low $p_{\mathrm{T}}$ region,
 in small centrality events,
 and at midrapidity,
 including the mass dependence of hadrons
 (for recent reviews, see Ref.~\cite{Huovinen:2003fa}).
This suggests that
 hydrodynamics could be reliable
for the description of the time evolution of soft sector
of matter produced in high energy heavy ion collisions at RHIC.
Certainly, it is more desirable to describe the time evolution of
the whole stage in high
energy heavy ion collisions by simulating collisions of initial nuclear
wave functions.
Instead, they simply assume that the system
created in heavy ion collisions reaches local thermal
equilibrium state
at some time.

Due to the above two reasons,
a model which treats a soft sector by hydrodynamics
and a hard sector based on a pQCD parton model
is turned out to be useful
 in order to understand experimental data at RHIC
 \textit{from low to high} $p_{\mathrm{T}}$.
Indeed, first attempts based on this concept has been done
by pQCD calculations which include hydrodynamic features
~\cite{Wang:2001cy,Gyulassy:2001kr,Gyulassy:2000gk}.
Motivated by these works,
we have recently developed a two component \textit{dynamical} model
 (hydro+jet model)~\cite{HIRANONARA}
 with a fully three dimensional hydrodynamic model~\cite{Hirano:2001eu}
  for the soft sector
 and pQCD jets for the hard sector
      which are computed via the PYTHIA code~\cite{pythia}.

Usually, it is possible to fit hadron spectra up to high momentum,
say $p_{\mathrm{T}} \sim$ 2-3 GeV/$c$, within hydrodynamics by adjusting
kinetic freeze-out temperature $T^{\mathrm{th}}$
which is a free parameter in the model \cite{Heinz:2002un}. 
Thus it is unclear which value of $T^{\mathrm{th}}$
 should be used when one wants to
add jet components into hydrodynamic components
 for the description of high $p_{\mathrm{T}}$ part.
However, we are free from this problem
thanks to inclusion of the early chemical freeze-out picture
into hydrodynamics.
One of the authors studied the effects
   of chemical freeze-out temperature $T^{\mathrm{ch}}$ 
which is separate from kinetic one $T^{\mathrm{th}}$
 in hydrodynamic model in Ref.~\cite{HiranoTsuda}.
It was found that the $p_{\mathrm{T}}$ slope for pions
remains invariant under the variation of $T^{\mathrm{th}}$
and that the hydrodynamic model with early chemical freeze out is able
to fit the transverse momentum distribution of pions up to 1-2 GeV/$c$.
Therefore, it is certain to incorporate hard partons
into the hydrodynamics with early freeze out
  in order to account for the high transverse momentum part of
  the hadronic spectrum.
We note that,
since we do not assume thermalization for the high $p_{\mathrm{T}}$ jets,
a hydrodynamical calculation with the initial conditions
taken from
pQCD+final state saturation model
\cite{Eskola:2001bf}
is different from ours.

In this paper, we shall study identified hadron spectra
from low to high $p_{\mathrm{T}}$ within the hydro+jet model.
In particular, 
  we focus on the influence of the hydrodynamic radial flow on the
 pQCD predictions for the transverse spectra.
Parameters in the hydrodynamic part of the model
have been already fixed
by fitting the pseudorapidity distribution.
Parameters related to the propagation
of partons are also obtained by fitting
  the neutral pion suppression factor by PHENIX
  and are found to be 
  consistent~\cite{HIRANONARA3}
   with the back-to-back correlation data from STAR~\cite{star:btob}.

The paper is organized as follows.
In Sec.~\ref{section:model},
   we describe the main features of our model.
We will represent results of transverse momentum distributions
 for pions, kaons, and protons
  in Sec.~\ref{sec:res:pt}.
Nuclear modification factor (suppression factor)
for identified hadrons and particle ratios are discussed
  in Sec~\ref{sec:res:raa}.
Elliptic flow for identified hadrons is discussed
  in Sec~\ref{sec:res:v2}.
Section~\ref{sec:summary} summarizes this paper.

\section{Model description} \label{section:model}

In this section, 
we explain in some detail the hydro+jet model
as a dynamical model to describe relativistic heavy ion collisions.

\subsection{Hydrodynamics}

Let us start with the review of our hydrodynamics.
Main features of the hydrodynamic part in the hydro+jet model
are the following.

Although initial conditions and pre-thermalization stages
are very important subjects in the physics of heavy ion collisions
(see, e.g., Ref.~\cite{Iancu:2003xm,dtson}),
these are beyond the scope of this paper.
Instead, assuming local thermal equilibrium of partonic/hadronic matter
at an initial time $\tau_0$,
 we describe afterward the space-time evolution of thermalized matter
 by solving the equations for energy-momentum conservation
\begin{equation}
\label{eq:ene-mom}
  \partial_{\mu}T^{\mu\nu}=0, \quad T^{\mu\nu}=(e+P)u^\mu u^\nu -Pg^{\mu\nu}
\end{equation}
in the \textit{full} three-dimensional
 Bjorken coordinate $(\tau,x,y,\eta_{\mathrm{s}})$.
Here 
$e$, $P$, and $u^\mu$ are, respectively, energy density, pressure,
and local four velocity.
$\tau=\sqrt{t^2-z^2}$ is the proper time
 and $\eta_{\mathrm{s}}=(1/2)\ln[(t+z)/(t-z)]$ is the space-time rapidity.
 Throughout this paper,
we consider baryon free matter $n_{\mathrm{B}}=0$ at RHIC energies.
In order to obtain reliable
   solutions of Eq.~(\ref{eq:ene-mom})
    especially in the longitudinal direction
    at collider energies,
$\tau$ and $\eta_{\mathrm{s}}$ are substantial choices
for time and longitudinal directions
rather than the Cartesian coordinate.

Assuming $N_{\mathrm{f}}=3$ massless partonic gas for the QGP phase,
an ideal gas EOS with a bag constant $B^{1/4}=247$ MeV is used in the
high temperature phase.
We use a hadronic resonance gas model with all hadrons up to $\Delta(1232)$
     for later stages of collisions. 
Possible finite baryonic effects
 such as a repulsive mean field~\cite{Sollfrank:1996hd}
 are not included because of
 the low baryon density at RHIC~\cite{Bearden:fw}.
Phase transition temperature is set to be $T_{\mathrm{c}}=170$ MeV.
For the hadronic phase,
a partial chemical equilibrium (PCE) model
with chemical freeze-out temperature $T^{\mathrm{ch}}=170$ MeV
is employed
to describe the early chemical freeze-out picture
of hadronic matter.
Although chemical freeze-out temperature
$T^{\mathrm{ch}} (\sim$ 160-170 MeV)
is usually found to be larger than
 kinetic freeze-out temperature $T^{\mathrm{th}}(\sim$ 100-140 MeV)
 from statistical model analyses and thermal model fitting~\cite{Shuryak:1999zh}, 
the sequential freeze out is not considered so far
in the conventional hydrodynamics except for a few work
\cite{HiranoTsuda,Arbex:2001vx,Teaney:2002aj,Kolb:2002ve}.
As a consequence of this improvement,
   the hadron phase cools down more rapidly
than the one in usual hydrodynamic calculations in which 
 $T^ {\mathrm{ch}}=T^ {\mathrm{th}}$ is
  assumed~\cite{HiranoTsuda,Arbex:2001vx}.
It should be emphasized that 
  the slope of pions in the transverse momentum distribution
  becomes insensitive to
 the choice of the kinetic freeze-out temperature $T^{\mathrm{th}}$
 and that the hydrodynamics with early chemical freeze out
reproduces the RHIC data of the pion transverse momentum
only up to 1.5 GeV/$c$~\cite{HIRANONARA2}.
This is one of the strong motivations
which leads us to combine our hydrodynamics
with non-thermalized hard components.

From hydrodynamic simulations, we evaluate hadronic spectra
which originate from thermalized hadronic matter.
For hadrons directly emitted from freeze-out hypersurface $\Sigma$,
we calculate spectra through
the Cooper-Frye formula \cite{Cooper:1974mv}
\begin{eqnarray}
\label{eq:CF}
E\frac{dN_i}{d^3p} = \frac{d_i}{(2\pi)^3}\int_\Sigma
\frac{p^\mu d\sigma_\mu}{\exp[(p^\mu u_\mu -\mu_i)/T^{\mathrm{th}}]\mp1},
\end{eqnarray}
where $d_i$ is a degeneracy factor, $\mu_i$ is a chemical potential, 
$p^\mu$ is a four momentum in the center of mass frame of colliding two nuclei,
and $-(+)$ sign is taken for bosons (fermions).
We should note the existence of chemical potentials $\mu_i$ for \textit{all}
hadrons under consideration
due to early chemical freeze out.
Typical values at $T^{\mathrm{th}}=100$ MeV are as follows:
$\mu_\pi = 83$ MeV, $\mu_K = 181$ MeV, and $\mu_p =\mu_{\bar{p}} = 349$ MeV.
For hadrons from resonance decays,
we use Eq.~(\ref{eq:CF}) for resonance particles at freeze out
and afterward take account of decay kinematics.
Here these resonances also have their own chemical potentials at freeze out.
We call the sum of the above spectra
the soft component or the hydro component throughout this paper.

Initial energy density at $\tau_0=0.6$ fm/$c$
 is assumed to be factorized
\begin{equation}
    e(x,y,\eta_{\mathrm{s}};b) = e_{\mathrm{max}}W(x,y;b)H(\eta_{\mathrm{s}}).
\end{equation}
 Here the transverse profile $W(x,y;b)$
   is proportional to the number of binary collisions
   and normalized as $W(0,0;0) = 1$,
  whereas longitudinal profile $H(\eta_{\mathrm{s}})$ is flat and unity near
midrapidity and falls off smoothly at large rapidity.
In $H(\eta_{\mathrm{s}})$, we have two adjustable parameters
$\eta_\mathrm{flat}$ and $\eta_\mathrm{Gauss}$ which
parametrize the length of flat region near midrapidity
and the width of Gaussian in the forward/backward rapidity region,
respectively. These parameters are chosen so as to reproduce the shape
of $dN/d\eta$ or $dN/dY$.

\begin{figure}[t]
\includegraphics[width=3.3in]{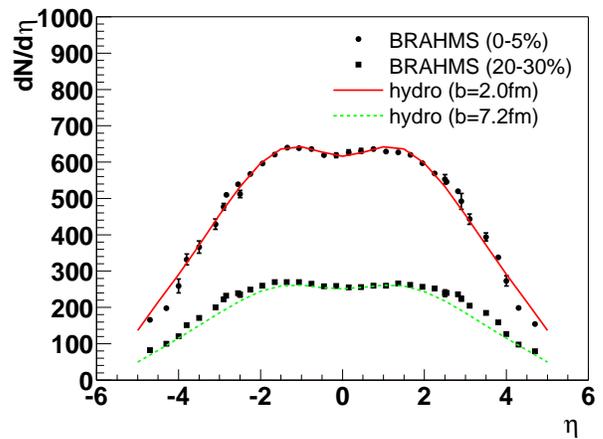}
\caption{(Color online)
Pseudorapidity distribution of charged particles
 in Au + Au  collisions at $\sqrt{s_{NN}}=200$ GeV
is compared to data from BRAHMS~\cite{Bearden:2001qq}.
Solid (dashed) line represents the hydrodynamic result
at b=2.0 (7.2) fm.
}
\label{fig:dndeta_hydro}
\end{figure}

We choose $e_{\text{max}}=40$ GeV/fm$^3$, 
$\eta_\mathrm{flat}=4.0$, and $\eta_\mathrm{Gauss}=0.8$.
As shown in Fig.~\ref{fig:dndeta_hydro}, the pseudorapidity distribution
of charged hadrons in 5\% central collisions
 observed by the BRAHMS Collaboration~\cite{Bearden:2001qq}
 is satisfactory reproduced
 by using the above parameters. Here we choose an impact parameter
as $b=2$ fm for this centrality.
These initial parameters
give us an average initial energy density about 5 GeV/fm$^3$
in the transverse plane $\eta_{\mathrm{s}}=0$ at $\tau = 1$ fm/$c$
\cite{Huovinen:2002rn}.
A contribution from minijets is neglected in the hydrodynamic fitting,
since it is less than 5\%
effect to the total hadron yield at RHIC
when we define minijets as particles with transverse momentum
larger than 2 GeV/$c$.
Initial conditions for transverse profile
are scaled by the number of binary
collisions. 
It is found that the 20-30\%
 semicentral collision data is also reproduced
 simply by choosing $b$ as 7.2 fm in the transverse profile $W$ \cite{noteonBCscaling}.

\begin{figure}[t]
\includegraphics[width=3.3in]{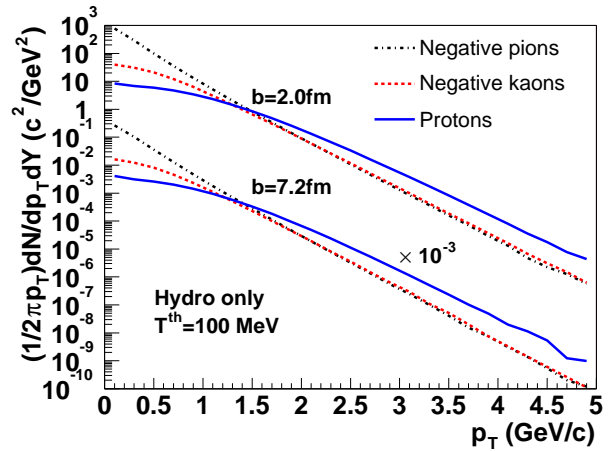}
\caption{(Color online)
Transverse momentum spectra for negative pions, negative kaons,
and protons from the hydro model with early chemical freeze out
in Au+Au collisions at $\sqrt{s_{NN}}=200$ GeV.
We choose an impact parameter as $b=2.0$ (7.2) fm
corresponding to 0-5\% (20-30\%) centrality.
Yields are divided by 10$^3$ for $b=7.2$ fm results.
}
\label{fig:hydro}
\end{figure}

In Fig.~\ref{fig:hydro},
we show the transverse spectra for negative pions, 
negative kaons, and protons
 in Au+Au collisions at $\sqrt{s_{NN}}=200$ GeV
  from the hydrodynamic model
  for impact parameters $b=2.0$ fm and 7.2 fm.
Thermal freeze-out temperature $T^{\mathrm{th}}=100$ MeV
is used in the calculation.
This choice is consistent with the data
at $\sqrt{s_{NN}}=130$ GeV~\cite{HiranoTsuda}.
The flatter behavior at low $p_{\mathrm{T}}$ for kaons and protons
is indeed a consequence of the radial flow effect.
A remarkable feature on the hydrodynamical result is that
 $p/\pi^{-} >1$ and $K^{-}/\pi^{-} \sim 1$
 above $p_{\mathrm{T}} \sim 2$ GeV/$c$.
It is, however, questionable to assume thermalization
at high $p_{\mathrm{T}}$ region.
In fact, hydrodynamical predictions overestimate elliptic flow data
  at the large transverse momentum region.
It is interesting to ask
at which $p_{\mathrm{T}}$ hydrodynamic behavior ceases
and switches to pQCD results.
We will see in the next section
how these hydrodynamical results are modified
by including the pQCD hard component.

\subsection{Jet propagations}

For the hard part of the model,
  we generate hard partons according to a pQCD
parton model. The number of jets at an impact parameter $\bm{b}$
are calculated from
\begin{equation}
N_{\mathrm{hard}}(\bm{b})= \int d^2\bm{r}_{\perp}\sigma_{\text{jet}}
T_A(\bm{r}_{\perp}-\bm{b}/2)T_B(\bm{r}_{\perp}+\bm{b}/2),
\end{equation}
 where $\sigma_{\text{jet}}$
is a hard cross section from leading order pQCD
convoluted by the parton distribution functions and
 multiplied by
a $K$-factor which takes into account higher order contributions.
$T_A$ and $\bm{r}_{\perp}$ are, respectively,
a nuclear thickness function
 normalized to be $\int d^2\bm{r}_{\perp} T_A = A$
 and
 a transverse coordinate vector.
Here we use the Woods-Saxon distribution for the nuclear density profile.
 We use PYTHIA 6.2~\cite{pythia} for the generation
of momentum spectrum of jets through
  $2\to2$ QCD hard processes.
  Initial and final state radiations are used
  to take into account the enhancement of higher-order contributions
  associated with multiple small-angle parton emission.

Scale $Q^2$ dependent nuclear shadowing effect is included
for the mass number $A$ nucleus
assuming the impact parameter dependence~\cite{Emel'yanov:1999bn}:
\begin{equation}
 S(A,x,Q^2,\bm{r}_{\perp}) = 1 + [S(A,x,Q^2)-1]
           \frac{AT_A(\bm{r}_{\perp})}{\int d^2 \bm{r}_{\perp}
           T_A(\bm{r}_{\perp})^2},
\end{equation}
where the EKS98 parametrization~\cite{eks98} is used for $S(A,x,Q^2)$.
Then the nuclear parton distribution function in this model has
the form
\begin{eqnarray}
f_A(A,x,Q^2,\bm{r}_{\perp}) &=& S(A,x,Q^2,\bm{r}_{\perp}) \nonumber\\
&\times&
    \left[\frac{Z}{A}f_p(x,Q^2)+\frac{(A-Z)}{A}f_n(x,Q^2) \right],\nonumber\\
\end{eqnarray}
where $f_p(x,Q^2)$ and $f_n(x,Q^2)$ are the
parton distribution functions for protons and neutrons.
We simply assume
the charge of a nucleus to be $Z=A/2$ in consistency 
with the soft part, since our fluids are
assumed to be isospin symmetric as well as baryon free matter.

Cronin effect~\cite{Cronin},
which has also been discovered in recent RHIC experiments~\cite{rhic:dA},
is usually considered as the multiple initial state scattering effect.
Understanding this effect becomes an important subject 
in RHIC physics~\cite{XNWang:1998ww,adjjm,pa}.
We employ the model in Ref.~\cite{XNWang:1998ww}
to take into account the multiple initial state scatterings,
in which initial $k_{T}$ is broadened proportional to the 
number of scatterings:
\begin{equation}
  \langle k^2_{T}\rangle_{NA} = \langle k^2_{T}\rangle_{NN}
    + \delta^2(Q^2)\left(
     \sigma_{NN}T_A\left(\bm{r}_{\perp}\right)-1
          \right),
\end{equation}
where 
$\sigma_{NN}$ is the inelastic nucleon-nucleon cross section
and
$\delta^2(Q^2)$ is the scale dependent $k_{T}$ broadening
per nucleon-nucleon collision whose explicit form can be found
in Ref.~\cite{XNWang:1998ww}.

We need to specify a scale which separates a soft sector from a hard sector,
in other words, a thermalized part from a non-thermalized part
in our model.
We include minijets with transverse momentum $p_{\mathrm{T,jet}}$
larger than 2 GeV/$c$
just after hard scatterings
in the simulation.
These minijets explicitly propagate through fluid elements.

\begin{figure}[t]
\includegraphics[width=3.3in]{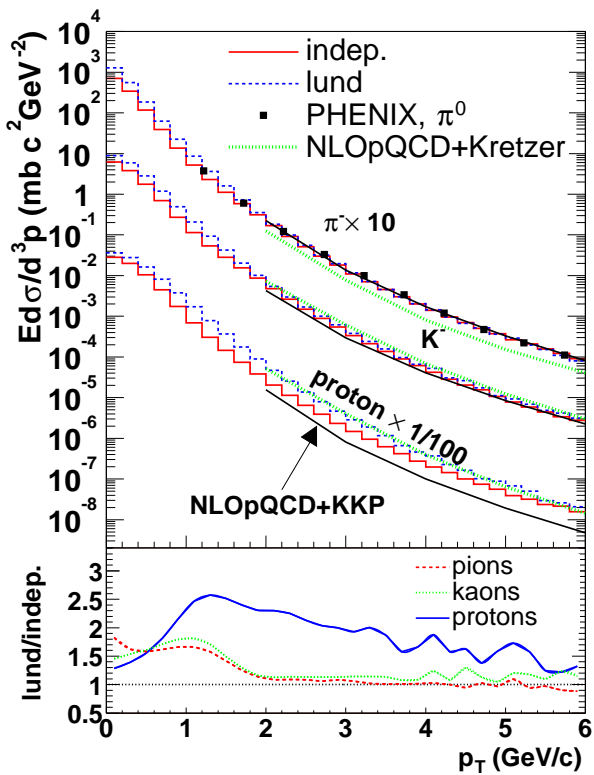}
\caption{(Color online)
Comparison with various models for inclusive pion, kaon,
and proton transverse momentum distributions in $pp$ collisions
at $\sqrt{s}=200$ GeV.
Solid and dotted histograms correspond to the results from
PYTHIA with independent fragmentation and
default Lund fragmentation respectively.
Solid and dotted lines are, respectively, from NLOpQCD calculations with
KKP and Kretzer fragmentation functions.
}
\label{fig:lund_vs_indep_vs_NLO}
\end{figure}

  Since we only pick up high $p_{\mathrm{T}}$
  partons from PYTHIA and throw them into fluids,
  there is ambiguity to connect color flow among partons.
Thus we use an independent fragmentation model option in PYTHIA
  to convert hard parton to hadrons instead of using the default
  Lund string fragmentation model.
We note that the independent fragmentation model should not be applied
at low transverse momentum region.
We have checked that the neutral pion transverse
  spectrum in $pp$ collisions at RHIC~\cite{Adler:2003pb} is well reproduced
  by selecting the $K$-factor $K=2.5$,
  the scale $Q=p_{\mathrm{T,jet}}/2$
  in the CTEQ5 leading order parton distribution
  function~\cite{cteq5},
  and the primordial transverse momentum
  $\langle k_{\mathrm{T}}^2\rangle_{NN}=1$ GeV$^2$/$c^2$
  as shown in Fig~\ref{fig:lund_vs_indep_vs_NLO}.
As shown in the bottom panel of the Fig.~\ref{fig:lund_vs_indep_vs_NLO},
independent fragmentation model predictions for pions and kaons 
are very close to those from the Lund string fragmentation model in
$p_{\mathrm{T}} > 2$ GeV/$c$,
where $K=2$ and $Q=p_{\mathrm{T,jet}}/2$ is used
in the Lund string model case and non-perturbative
inelastic soft processes are included.
However, the yield of protons from the independent fragmentation scheme
 becomes much less than that from Lund string model
 predictions
 as seen in Fig.~\ref{fig:lund_vs_indep_vs_NLO}.
We found that the Lund fragmentation scheme is favored
in terms of the recent STAR data of protons
in $pp$ collisions~\cite{Adams:2003qm}.
In what follows, we make corrections for our
 $p_{\mathrm{T}}$
spectra of kaons and protons
in Au+Au collisions according to the result in
the bottom panel of Fig.~\ref{fig:lund_vs_indep_vs_NLO}.

In order to see the theoretical uncertainties
on the fragmentation scheme deeply,
we also plot the results from NLOpQCD calculations~\cite{incnlo}
with the MRST99~\cite{mrst99} set of parton distribution functions.
In Fig.~\ref{fig:lund_vs_indep_vs_NLO},
 we show results from two different fragmentation functions.
The solid lines are obtained from KKP fragmentation functions~\cite{kkp}
with renormalization scale $\mu$, factorization scale $M$,
and fragmentation scale $M_f$ equal to $p_{\mathrm{T}}$.
NLOpQCD prediction with KKP fragmentation functions
 is consistent with the pion data.
NLOpQCD predictions
with the Kretzer fragmentation functions~\cite{Kretzer}
assuming $\mu=M=M_F=p_{\mathrm{T}}/2$
underestimate pion yields, while yields for kaons and protons
are the same as the predictions from
the PYTHIA default Lund string fragmentation model.

%
%

Initial transverse positions of jets at an impact parameter $\bm{b}$
 are determined randomly
 according to the probability $P(\bm{r}_{\perp},\bm{b})$ specified by
 the number of binary collision distribution,
\begin{equation}
   P(\bm{r}_{\perp},\bm{b})
    \propto T_A(\bm{r}_{\perp}+\bm{b}/2)T_A(\bm{r}_{\perp}-\bm{b}/2).
\end{equation}
Initial longitudinal position of a parton
  is approximated by the boost invariant distribution~\cite{Bjorken:1982qr}:
  $\eta_{\mathrm{s}}=Y$,
 where
$Y=(1/2)\ln[(E+p_z)/(E-p_z)]$ is the rapidity of a parton.
Jets are freely propagated up to the initial time $\tau_0$ of hydrodynamic
simulations
by neglecting the possible interactions in the pre-thermalization stages.
Jets are assumed to travel with straight line trajectory
in a time step:
\begin{eqnarray}
 \Delta r_i &=& \frac{p_i}{m_{\mathrm{T}}\cosh(Y-\eta_{\mathrm{s}})}
 \Delta \tau, \quad (i=x,y),\\
 \Delta\eta_{\mathrm{s}} &=& \frac{1}{\tau}\tanh(Y-\eta_{\mathrm{s}})
  \Delta \tau,
\end{eqnarray}
where $m_{\mathrm{T}}=\sqrt{m^2+\bm{p}_{\mathrm{T}}^2}$ is a transverse mass.

Jets can suffer interaction with fluids and lose their energies.
We employ the approximate first order formula (GLV formula)
 in opacity expansion
 from the reaction operator approach~\cite{Levai}
 for the energy loss of partons throughout this work.
The opacity expansion 
is relevant for the realistic heavy ion reactions
where the number of jet scatterings is small.
The energy loss formula for coherent scatterings
in matter
has been applied to analyses of
heavy ion reactions
taking into account the expansion of the system
 \cite{Gyulassy:2000gk,Gyulassy:2001kr,Wang:2001cy,VG}.
The approximate first order formula in this approach can be written as
\begin{equation}
\Delta E = C \int_{\tau_0}^{\infty} d\tau
\rho\left(\tau, \bm{x}\left(\tau\right)\right)
(\tau-\tau_0)\ln\left({\frac{2E_0}{\mu^2 L}}\right).
\label{eq:GLV}
\end{equation}
Here $C$ is an adjustable parameter and
$\rho(\tau,\bm{x})$ is a thermalized parton density
 in the local rest frame of fluid elements in the hydro+jet approach
\cite{noteweb}.
$\bm{x}\left(\tau\right)$ and $E_0$ are the position and
the initial energy of a jet, respectively.
The initial energy $E_0$ in Eq.~(\ref{eq:GLV}) 
is Lorentz-boosted by the flow velocity and replaced by $p_0^{\mu} u_{\mu}$
where $p_0^\mu$ is the initial four momentum of a jet and $u_{\mu}$ is a
local fluid velocity.
We take a typical screening scale $\mu = 0.5$ GeV
and effective path length $L=3$ fm which is chosen from the lifetime
of the QGP phase.
Here we choose $C=0.45$~\cite{commentC}
 which is found to reproduce the neutral pion $R_{AA}$
 defined by Eq.~(\ref{eq:raa})~\cite{phenix:pi0_200}.
Our purpose here is not a detailed study of jet quenching mechanisms.
Instead, we first fit the suppression factor for neutral pions
and next see other hadronic spectra.

Feedback of the energy to fluid elements
in central collisions was found to be
about 2\% of the total fluid energy.
Hence we can safely neglect its effect on hydrodynamic evolution
in the case of the appropriate amount of energy loss.

%
\begin{figure}[t]
\includegraphics[width=3.3in]{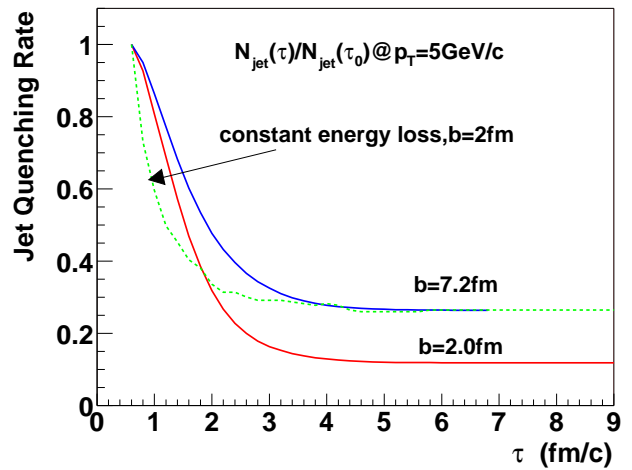}
\caption{(Color online)
Jet quenching rate
$N_{\text{jet}}(\tau)/N_{\text{jet}}(\tau_0)$
for $p_{\mathrm{T}} = 5$ GeV/$c$ jets
in Au+Au collisions at $\sqrt{s_{NN}}=200$ GeV.
Jet quenching rate for 10 GeV/$c$ jets is very similar to 
that of the 5 GeV/$c$ jets.
}
\label{fig:jetq}
\end{figure}
%

In Fig.~\ref{fig:jetq}, we show the jet quenching rate
as a function of proper time
for 5 GeV/$c$ jets.
We count the number of partons with $4.5 < p_{\mathrm{T,jet}} < 5.5$ GeV/$c$
at each time step, and then
define the ratio of the current number of jets 
to the initial number of jets
$N_{\text{jet}}(\tau)/N_{\text{jet}}(\tau_0)$.
Most jet quenching is completed at early times less than 4 fm/$c$.
For comparison, we also plot the jet quenching rate for a constant
energy loss case $dE/dx \propto \rho(\tau)$.
Jet quenching is almost finished at $\tau \sim 2$ fm/$c$
in the case of constant energy loss.
From Fig.~\ref{fig:jetq}, the degree of decrease for the
jet quenching rate in the GLV formula becomes milder
and continues longer than that in the incoherent model.
This is due to the existence of $\tau$ in the integrand in Eq.~(\ref{eq:GLV})
which comes from the property of coherent
(Landau-Pomeranchuk-Migdal~\cite{LPM}) effect. 
Contrary to the simple Bjorken's ansatz~\cite{Bjorken:1982qr},
$\rho(\tau) =\rho_0 \tau_0/\tau$,
there exists transverse flow and the parton density profile
in the transverse plane is not flat in our simulations. 
This is the reason why jets are quenched only in the QGP phase
and why jet quenching in the mixed phase is totally negligible.

We include $p_{\perp}$ broadening
accompanied by the energy loss of jets
with the formula
$\langle p^2_{\perp}\rangle \sim \int d\tau\rho(\bm{r})$
as in Ref.~\cite{HIRANONARA3}.
We found that this effect is small in all results
in this paper.

Within our model, we neglect energy loss before thermalization,
in our case, $\tau < 0.6$ fm/$c$.
One would ask if it is important to take into account
the energy loss effects before thermalization
 because parton density has the maximum value.
We can, however, fit the suppression factor $R_{AA}$ by rescaling the
energy loss parameter $C$ when the initial time $\tau_0$ is changed.
The question about the jet quenching before thermalization is
beyond our model description.
As a possible model for 
a study of jet interactions at early times,
 propagation of jets in the  classical Yang-Mills fields
 based on the idea 
 of the Color Glass Condensate~\cite{MV,Iancu:2003xm}
 is proposed in Ref.~\cite{Shuryak:2002ai}.
It would be interesting to take a numerical results
from the full lattice calculations~\cite{KNV} for the calculations
of jet energy loss at the very early stages of the collisions.

\section{Results}
  We discuss in this section transverse dynamics for
pions, kaons, and protons from the hydro+jet model
focusing on the intermediate $p_{\mathrm{T}}$ where interplay between
soft and hard components is expected to be crucial.
As mentioned in the previous section,
a parameter for jet quenching $C$ was already fixed by fitting
the observed data for neutral pions in central
Au+Au collisions from PHENIX.
Freeze-out temperature $T^ {\mathrm{th}}=100$ MeV is used for
hydrodynamics.
All results in this section
 are for midrapidity $\mid \eta \mid < 0.35$.

\subsection{Transverse momentum distributions for identified particles}
 \label{sec:res:pt}

%
\begin{figure}[t]
\includegraphics[width=3.3in]{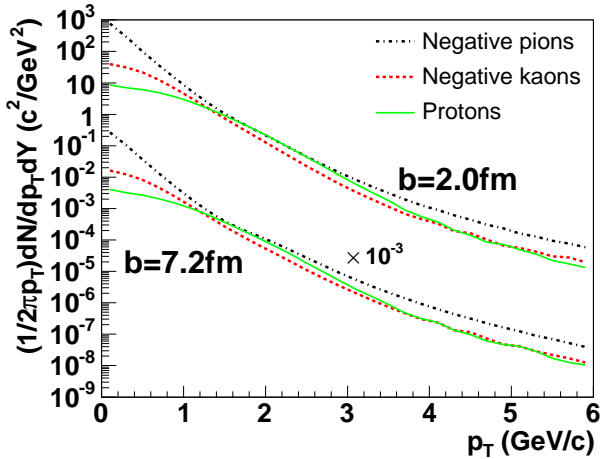}
\caption{(Color online)
Transverse momentum spectra for negative pions, negative kaons,
and protons from the hydro+jet model
in Au+Au collisions at $\sqrt{s_{NN}}=200$ GeV
at the impact parameter
  of $b=2.0$ and 7.2 fm. 
Yields are divided by 10$^3$ for $b=7.2$ fm results.
}
\label{fig:hydrojet}
\end{figure}
%

First, we show the transverse momentum distributions
 for pions, kaons, and protons
 from the hydro+jet model in Fig.~\ref{fig:hydrojet}
 in central as well as semicentral Au+Au collisions at RHIC.
Each spectrum is the sum of the soft component and the hard component.
Before summation, the hard component is multiplied
by a ``switch" function~\cite{Gyulassy:2000gk}
$\{1+\tanh[2(p_{\mathrm{T}}-p_{\mathrm{T,cut}})]\}/2$
(where $p_{\mathrm{T}}$ is in the unit of GeV/$c$ and
$p_{\mathrm{T,cut}}$ = 2 GeV/$c$)
in order to cut the unreliable components
from the independent fragmentation scheme
and also to fit $R_{AA}$ for neutral pions~\cite{phenix:pi0_200}.
We have checked the cutoff parameter dependence
in the switch function
on the pion spectrum
and found that we are not able to fit
the pion data anymore even with $p_{\mathrm{T,cut}} = 1.8$ or 2.2 GeV/$c$.
 So the ambiguity of the cutoff can be removed to fit
the pion data within our approach.

At low transverse momentum region $p_{\mathrm{T}}< 1$ GeV/$c$,
  the shapes remain the same as hydro predictions as one can
 check from Fig.~\ref{fig:hydro}.
Also at high transverse momentum,
    spectra are identical to those of pQCD predictions
    with an appropriate amount of jet quenching.

Our calculation includes interactions of minijets with QGP fluids.
We also note that 
 there remains a  pQCD like power law behavior
 in all hadrons
at high transverse momentum.
This may indicate no hint for the thermalization at high transverse momentum.
However, energy loss results in a parallel shift of hadronic spectra,
since the
energy loss model used in this paper shows almost flat
quenching pattern as shown in our previous analysis~\cite{HIRANONARA3}.

%
%
%
\begin{figure}[t]
\includegraphics[width=3.3in]{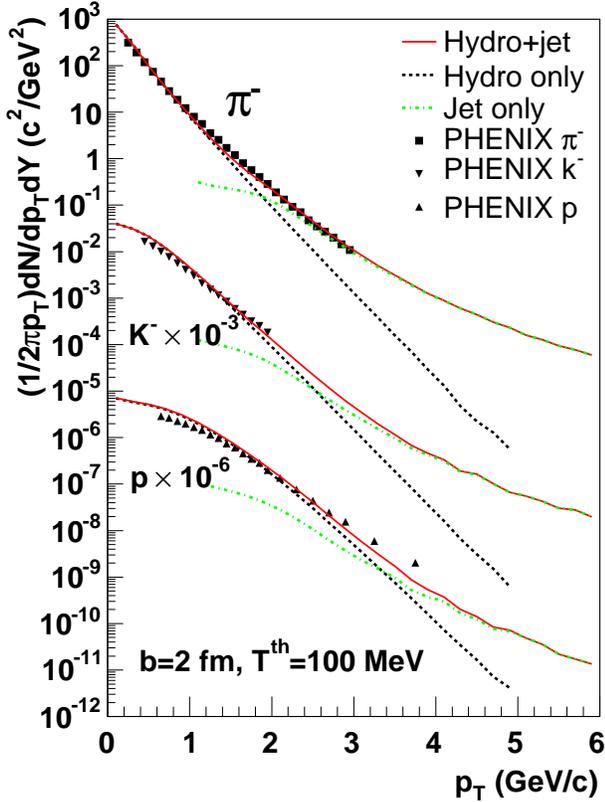}
\caption{(Color online)
Each contribution from hydrodynamics and minijets for $\pi^-$,
$K^-$, and $p$
in Au+Au collisions at $\sqrt{s_{NN}}=200$ GeV
at the impact parameter of $b=2.0$ fm.
Yield of negative kaons (protons) is divided by 10$^3$ (10$^6$).
PHENIX data are from Ref.~\cite{Adler:2003cb}.
}
\label{fig:pikp}
\end{figure}
%

In Fig.~\ref{fig:pikp}, 
  we decompose the spectra into hydro parts and minijet parts.
Here the yields from hard components are multiplied by the switch
function again.
It is seen that both soft and hard components
are important for the hadron spectra in the transverse momentum
of the range around $2 \lsim p_{\mathrm{T}} \lsim5$ GeV/$c$
depending on the hadron mass.
We can define the crossing point of transverse momentum
$p_{\mathrm{T,cross}}$ at which the yield from the soft part is identical
to that from the hard part.
$p_{\mathrm{T,cross}}$ moves toward high momentum with mass of particles
   because of the effects of radial flow.
In central collisions, $p_{\mathrm{T,cross}} \sim$ 1.8,
2.5, and 3.5 GeV/$c$ for pions, 
kaons, and protons, respectively.
Minijet spectra are recovered at $p_{\mathrm{T}}\sim 3.4$ GeV/$c$ for pions,
 $p_{\mathrm{T}}\sim 4.0$ GeV/$c$ for kaons,
 and $p_{\mathrm{T}}\sim 5.0$ GeV/$c$ for protons.

We give some remarks here:

(a) The point at which hydrodynamic and pQCD spectra cross is determined
by the dynamics of the system:
 The radial flow pushes the soft components toward high $p_{\mathrm{T}}$
 region, while the dense matter reduces the pQCD components through
 parton energy loss. The crossing of two spectra causes
  by interplay of these two effects.

(b) At $p_{\mathrm{T}}=2$-3 GeV/$c$, the yields of pions and kaons
are no longer occupied by soft hydrodynamic component. On the other hand,
the proton yield from pQCD prediction is about
ten times smaller than that of hydro
in the transverse momentum region.

(c) One may try to extract the strength of radial flow
and the kinetic freeze-out temperature from experimental data
through the hydrodynamics-motivated fitting model.
Then one should pay attention to the fitting range of the transverse
momentum. In particular, $p_{\mathrm{T}}$ spectrum for pions may have
no room to fit by a simple thermal spectrum: Contribution from resonance
decays becomes important below $p_{\mathrm{T}} \sim 0.5$ GeV/$c$,
while the hard component slides in the soft component
near $p_{\mathrm{T}} \sim 1.0$ GeV/$c$.

(d) We predict positions of the inflection point
where $p_{\mathrm{T}}$ spectrum becomes convex to concave:
$p_{\mathrm{T}} \sim 3$ GeV/$c$ for kaons and 
$\sim 4$ GeV/$c$ for protons. These are the indicators of
a transition from soft physics to hard physics. 

The amount of the hydrodynamic contributions
 to the hadron yields for each particle
 found in the hydro+jet model
 is very similar to that found in Ref.~\cite{Peitzmann} in which 
hybrid parametrization of hydrodynamics with the spectral shape
in $pp$ collisions.
It is also remarkable that
  baryon junction~\cite{Vitev:2002wh,ToporPop:2002gf}
  and
  quark coalescence models~\cite{Hwa:2002tu,Fries,Greco}
 predicts the same behavior.
Quark coalescence models are successful in explaining
 the mass dependence of $p_{\mathrm{T}}$ slopes~\cite{ALCOR,MICOR}.
For example,
one can easily understand the difference of the transverse slopes
of baryons and mesons from a quark coalescence hadronization mechanism.
A baryon momentum is a sum of three quarks
 (quark momenta must be almost parallel in order to cluster),
 but a momentum of mesons is a sum of two quarks.
It is interesting to see, for example, $\phi$ meson spectrum
in order to distinguish the mass effects in hydrodynamics
from meson-baryon effects in coalescence models.

\subsection{Suppression factors and particle ratios}
\label{sec:res:raa}

%
\begin{figure}[t]
\includegraphics[width=3.3in]{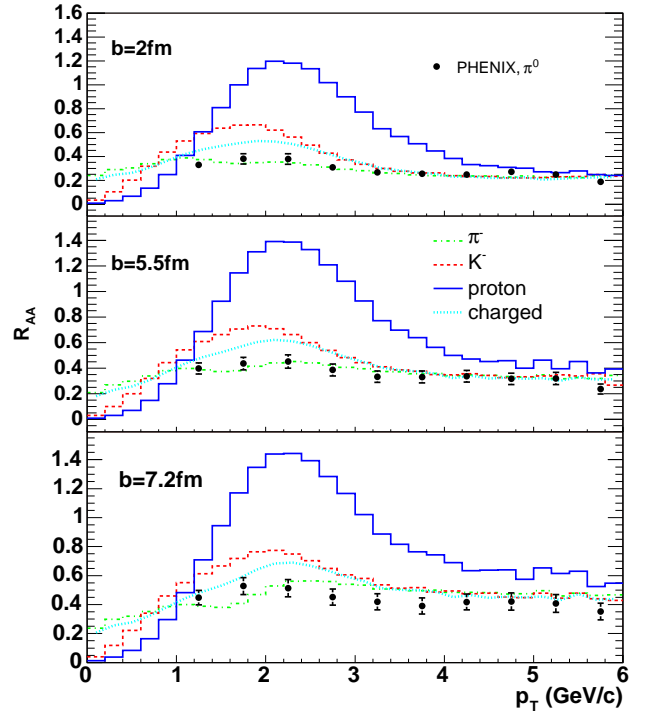}
\caption{(Color online)
Impact parameter dependence of the suppression factors $R_{AA}$
in Au+Au collisions at $\sqrt{s_{NN}}=200$ GeV
 as a function of $p_{\mathrm{T}}$ for $\pi^-$,
$K^-$, and $p$.
$R_{AA}$ for charged hadrons is also shown in dotted lines.
Experimental data of $R_{AA}$
for neutral pions~\cite{phenix:pi0_200}
is obtained by PHENIX.
For details, see text.
}
\label{fig:raa}
\end{figure}
%

We now turn to the study of the suppression factors $R_{AA}$
 for each hadron defined by
\begin{equation}
\label{eq:raa}
 R_{AA} = \frac{\frac{dN^{A+A}}{d^2p_{\mathrm{T}}dY}}
               {N_{\mathrm{coll}}\frac{dN^{p+p}}{d^2p_{\mathrm{T}}dY}}.
\end{equation}
It is very instructive to study $R_{AA}$ behaviors for identified hadrons
toward a comprehensive understanding of
intermediate transverse momentum region.

Figure~\ref{fig:raa} shows the suppression factors $R_{AA}$ for
pions, kaons, protons, and charged hadrons respectively
in Au+Au collisions at RHIC
for impact parameters $b=2.0$, 5.5, and 7.2 fm.
Our results for pions are compared with PHENIX
data~\cite{phenix:pi0_200}.
We use $pp$ spectra from Lund string model for protons and kaons
in the plots.
$R_{AA}$ for protons using the independent fragmentation
model becomes too large $\sim2.5$ at $p_{\mathrm{T}}\sim 2$ GeV/$c$.
This result simply comes from the fact that
the independent fragmentation model is inconsistent with
$pp$ data for protons as discussed in Sec.~\ref{section:model}.
Note that the numerator in Eq.~(\ref{eq:raa})
is almost free from the hard components at $p_{\mathrm{T}} < 3$ GeV/$c$ in proton case.
We find protons are not suppressed $R_{AA}>1$
at a momentum range of $1.5<p_{\mathrm{T}}<2.5$ GeV/$c$.
Pions, on the contrary, are largely suppressed for all momentum range.
Our calculations for protons become identical to
those of pQCD predictions at
a momentum above 5 GeV/$c$.
This is the same result as other model predictions~\cite{Fries,Greco,Peitzmann}.
In any case, these results
are easily understood from Fig.~\ref{fig:pikp}:
The crossing point $p_{\mathrm{T,cross}}$ depends on the hadronic species,
thus $R_{AA}$ only for pions reflects jet quenching effect,
while the larger value of $R_{AA}$ for protons simply
comes from radial flow, not the absence of jet quenching. 
We should mention that
   above $p_{\mathrm{T}}\sim 5$ GeV/$c$,
   suppression factors for identified hadrons
   converge to almost the same value.
It is also seen that the suppression factors
for kaons and protons
have almost no centrality dependence within this impact parameter range.

Recent data from PHENIX~\cite{Adler:2003kg} and STAR~\cite{STAR:Lambda_K}
for protons and $\Lambda$'s
show that
the nuclear modification factors
for  $p$, ${\bar p}$, and $\Lambda$ 
 in the $p_{T}$ range of $1.5<p_{T}<4.5$ GeV/$c$
 are almost constant.
However, our results of $R_{AA}$ for protons 
  seem to decrease to smaller value
 with transverse momentum
  faster than data.


From RHIC data~\cite{phenix:pi0_130,phenix:pi0_200,star:highpt,star:charged},
$R_{AA}$ for charged particles is larger
than the one for pions in moderate high $p_{\mathrm{T}}$ region.
In our model,
this also simply results from the average of the above three suppression
factors weighted by each yield (see also Fig.~\ref{fig:hydrojet})
as shown in Fig.~\ref{fig:raa} by the dotted lines.

%
\begin{figure}[t]
\includegraphics[width=3.3in]{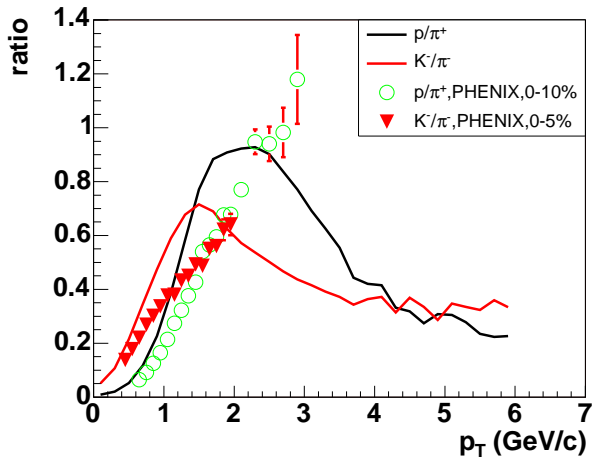}
\caption{(Color online)
Ratios of $N_{p}$ to $N_{\pi^-}$ and $N_{K^-}$ to $N_{\pi^-}$
as a function of $p_{\mathrm{T}}$
in Au+Au collisions at impact parameter $b=2$ fm.
PHENIX data~\cite{Adler:2003cb} are also ploted for comparsion.
}
\label{fig:ratio}
\end{figure}
%

We show in Fig.~\ref{fig:ratio}
 proton to negative pion ratio and negative kaon to negative pion
ratios as a function of the transverse momentum
in Au+Au collisions at $\sqrt{s_{NN}}=200$ GeV
for the impact parameter of $b=2$ fm
together with the PHENIX data~\cite{Adler:2003cb}.
Without depending on baryon junction mechanism or quark coalescence models,
we also obtain that $p/\pi^-$ ratio becomes close to 
unity due to the consequences
of hadron species dependent $p_{\mathrm{T,cross}}$.
Ratios become $p/\pi^- \sim 0.2$ and $K^-/\pi^- \sim 0.3$
 above $p_{\mathrm{T}} \sim 5$ GeV/$c$
 which are the consequences of pQCD predictions.
It should be noted that, 
if the baryonic and isospin chemical potentials are
included in the hydrodynamic simulation,
$p/\pi^-$ ratio can slightly be changed
in low $p_{\mathrm{T}}$ region: Baryon (isospin)
chemical potential pushes up (down) proton yield
from hydrodynamic components.

\subsection{Elliptic flow for identified particles}
  \label{sec:res:v2}

Azimuthal asymmetry for non-central heavy ion collisions
is generally considered to be generated only by the final state
interactions of matter created in the collisions.
In hydrodynamic models, elliptic flow is created
 by the anisotropic initial configuration of
 high pressure matter
 which might be the QGP phase.

\begin{figure}[t]
\includegraphics[width=3.3in]{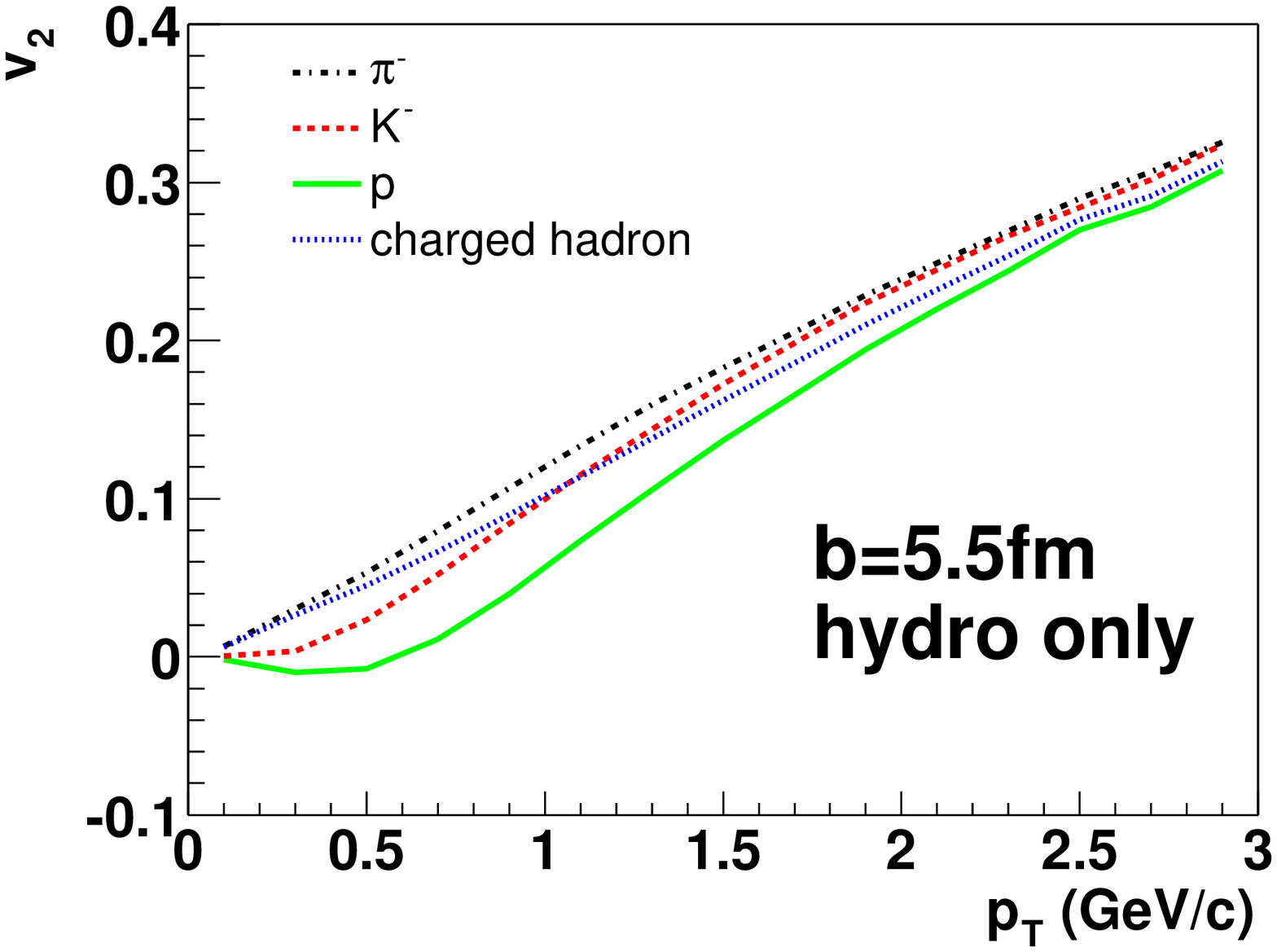}
\includegraphics[width=3.3in]{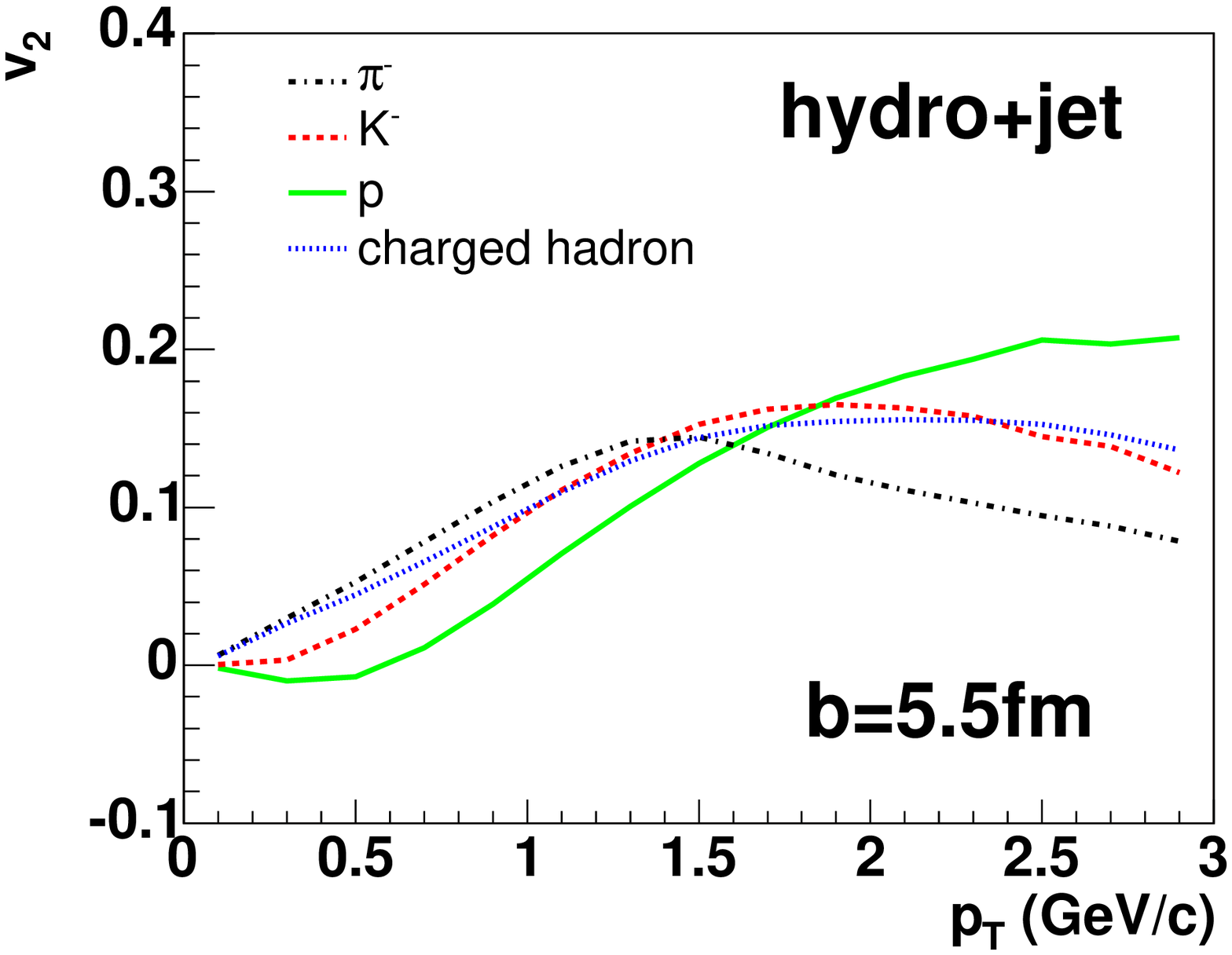}
\caption{(Color online)
Transverse momentum dependence of the elliptic flow
$v_2(p_{\mathrm{T}})$ for $\pi^-$, $K^-$, and $p$
in Au+Au collisions at $\sqrt{s_{NN}}=200$ GeV 
at impact parameter $b=5.5$ fm
from the hydrodynamic model (upper panel)
and the hydro+jet model (lower panel).
$v_2$ for charged hadrons are also represented in dotted lines.
}
\label{fig:v2}
\end{figure}

Hydrodynamic predictions on the transverse momentum dependence of
elliptic flow $v_2$ show almost linear increase for all particles.
However, the experimental data saturate at high $p_{\mathrm{T}}$
~\cite{STAR:Lambda_K,Adler:2002ct,Manly:2002uq}.
More interestingly, pion $v_2$ is larger than that of protons
at $p_{\mathrm{T}} < 1$ GeV/$c$,
while proton $v_2$ becomes larger than
pion $v_2$ at some $p_{\mathrm{T}}$~\cite{Adler:2003kt}.
 Hydrodynamic calculations are successful in reproducing
the mass dependence of the $v_2$ in the low transverse momentum
region~\cite{KHHH,Teaney:2000cw,Hirano:2001eu,HiranoTsuda}.
$v_2$ for pions are always greater than that of protons in hydrodynamics
 and, eventually, $v_2$ becomes almost mass independent at high transverse
momenta as shown in Fig.~\ref{fig:v2}.
On the other hand,
 to understand the observed azimuthal asymmetry
 at large $p_{\mathrm{T}}$,
it was showed that the jet interaction with matter generates
 the azimuthal asymmetry for non-central collisions
~\cite{Gyulassy:2000gk,Wang:2000fq,Shuryak:2001me}.

%
%

We demonstrate in Fig.~\ref{fig:v2} that,
 by combining minijet components
with hydrodynamics, pion $v_2$ can be reduced faster
 than proton $v_2$ at moderate high transverse momentum.
The hydro+jet predictions on $v_2$ for identified particles
in Au+Au collisions at RHIC for impact parameter $b=5.5$ fm
are compared to hydro results
 in Fig.~\ref{fig:v2}.
The magnitude of $v_2$ for kaons and protons
 becomes larger than $v_2$ for pions at about $p_{\mathrm{T}} > 1.3$ GeV/$c$.
The shape of $v_2$ for pions saturates faster than those of kaons and protons,
because the fraction of hydro components for pions are much smaller than
that for kaons or protons in this $p_{\mathrm{T}}$ region.
This is again the consequence of radial flow effect.
We demonstrate that
 the saturation point in transverse momentum
 depends on the hadron mass.
As a whole effect of the sum of pions, kaons, and protons,
the saturation point of $v_2$
for charged particles in transverse momentum is turned out
to be $p_{\mathrm{T}}=1.5$ GeV/$c$ in our model
at $b=5.5$ fm as one can read from Fig.~\ref{fig:v2}.

Our semi-macroscopic model produces consistent behavior in $v_2$
  with the experimental data from PHENIX~\cite{Adler:2003kt}.
Recently,
a microscopic description of quark coalescence model~\cite{Molnar:2003ff}
shows the crossing of
meson and baryon $v_2$'s at $p_{\mathrm{T}}\sim 1$ GeV/$c$.
In the simple coalescence model where  all partons have
similar elliptic flow, elliptic flow for baryons roughly
1.5 times stronger than for mesons.
On the other hand,
 our approach will have only mass dependence on the elliptic flow
indicating $v_{\Lambda}\sim v_{\phi}$.
Therefore, it is interesting to see, for example, $\phi$ meson
elliptic flow to clarify the origin of the elliptic flow.

We have studied $v_2$ in the momentum range where both soft and hard
contributions are important.
It is interesting to see $v_2$ up to 10 GeV/$c$. Experimental data
show that $v_2$ at high momentum saturates~\cite{Adler:2002ct}.
Systematic study on the elliptic flow within our model is under way
including centrality as well as rapidity dependence.

\section{Summary}\label{sec:summary}

We have studied  the interplay of soft and hard components
  by looking at
  $p_{\mathrm{T}}$ spectra,
  suppression factors,
  hadron ratios,
  and elliptic flow
  for identified particles 
  within the hydro+jet model.
By taking into account both hydrodynamic radial flow and quenched pQCD
spectra,
it was found that
$p_{\mathrm{T,cross}}$, at which the yield from the 
soft component is identical to the one from the hard component,
depends on the hadron species:
$p_{\mathrm{T,cross}} \sim 1.8, 2.5,$ and 3.5 GeV/$c$ for
pions, kaons, and protons in Au+Au central collisions at RHIC.
This difference comes from the interplay between the radial flow
for the soft part and the jet quenching for the hard part.
From the consequences of the interplay between soft and hard hadronic
components,
we showed $p/\pi^- \sim 1$ and $R_{AA}(p_{\mathrm{T}})>1$
 at intermediate $p_{\mathrm{T}}$ for protons.
We also showed that
 the mass dependence of the strength of $v_2(p_{\mathrm{T}})$
  in the intermediate $p_{\mathrm{T}}$ region
   is also explained by the radial flow + pQCD
 components.
Hydrodynamic radial flow plays an important role to understand
the transverse dynamics when hadron mass is large.

\begin{acknowledgements}
 The authors are grateful to S.~Kretzer for discussion about
fragmentation functions.
We also thank M.~Gyulassy for
giving us an opportunity to attend the visitor program at Columbia
University.
We acknowledge the hospitality of the nuclear theory groups at
Columbia University and the Institute for Nuclear Theory
where parts of this manuscript were completed.
The work of T.H.~is supported by
RIKEN.
Y.N.'s research is supported by the DOE under Contract No.
DE-FG03-93ER40792.
\end{acknowledgements}


\begin{thebibliography}{99}
\bibitem{noteonBCscaling}
It is often said that the multiplicity scales with
the number of participants, not binary collision.
This does not mean that the initial energy scales with
the number of participants, since the energy itself is not
directly proportional to the final multiplicity of hadrons.
 There is also so-called $pdV$ work during expansion,
hence $dE/dy$ is not a conserved number.
The behavior of centrality is consistent
with previous work. See, i.e., the last paper in Ref.~\cite{KHHH}.
\bibitem{commentC}
We used $C=0.35$ in the previous paper~\cite{HIRANONARA3} in which
neither nuclear shadowing effects
nor impact parameter dependent intrinsic $k_{\mathrm{T}}$
is included.
We note that
our results for back-to-back correlations are the same
after inclusion of nuclear shadowing and
impact parameter dependent intrinsic $k_{\mathrm{T}}$.
\end{thebibliography}
\end{document}